\documentclass[nofootinbib,showpacs,prd,superscriptaddress,twocolumn]{revtex4}

\usepackage{graphics,graphicx,rotating,color}
\usepackage[percent]{overpic}
\usepackage{mathrsfs,amsmath,amssymb}

\newcommand{\beq}{\begin{equation}}
\newcommand{\eeq}{\end{equation}}
\newcommand{\bea}{\begin{eqnarray}}
\newcommand{\eea}{\end{eqnarray}}

\begin{document}

\title{Bowen-York trumpet data and black-hole simulations}

\author{Mark Hannam} 
 \affiliation{Physics Department, University
  College Cork, Cork, Ireland} 
  
\author{Sascha Husa}
\affiliation{Departament de F\'isica, 
  Universitat de les Illes Balears, Cra.\ Valldemossa Km.\ 7.5, Palma 
de Mallorca, E-07122 Spain}
  
\author{Niall \'O~Murchadha} 
  \affiliation{Physics Department, University
  College Cork, Cork, Ireland}

\date{\today}

\begin{abstract}
The most popular method to construct initial data for black-hole-binary simulations
is the puncture method, in which compactified wormholes are given linear and
angular momentum via the Bowen-York extrinsic curvature. When these data are
evolved, they quickly approach a ``trumpet'' topology, suggesting that it would be
preferable to use data that are in trumpet form from the outset. To achieve this, 
we extend the puncture method to allow the construction of 
Bowen-York trumpets, including an outline of an existence and uniqueness proof of the
solutions. We construct boosted, spinning and binary Bowen-York
puncture trumpets using a single-domain pseudospectral elliptic solver, 
and evolve the binary data and compare with standard 
wormhole-data results. We also show that for boosted trumpets 
the black-hole mass can be prescribed {\it a priori}, 
without recourse to the iterative procedure that is 
necessary for wormhole data. 
\end{abstract}


\maketitle

\section{Introduction}

Numerical solutions of the full Einstein equations for the last orbits
and merger of compact binary systems are important for the developing
field of gravitational-wave astronomy. In the case of black-hole binaries,
long-term simulations became possible in 
2005~\cite{Pretorius:2005gq,Campanelli:2005dd,Baker:2005vv}, and within
the last few years the field has developed to the point where the 
gravitational-wave (GW) signal from such systems can 
be calculated to essentially the required accuracy of current 
GW detectors~\cite{Hannam:2009hh}, and work is underway to incorporate
these results into GW searches~\cite{Aylott:2009ya}. 
However, only a small fraction of the 
full black-hole-binary parameter space has yet been 
studied~\cite{Hannam:2009rd}, and its full exploration will require yet more
accurate and efficient numerical simulations. The first step in any 
simulation is the production of initial data, and these determine 
in part the accuracy and physical fidelity of the final simulation; 
that is the focus of this paper.

The 3+1 approach to solving Einstein's equations consists of specifying
initial data (the metric and its time derivative on one constant-time
slice of spacetime), and then evolving that data forward in time. Valid initial
data satisfy a set of constraint equations, and a given solution to the constraints
will represent a certain physical situation in a certain set of coordinates. We are
then faced with the problem of finding constraint-satisfying data that both represent
the physical situation we wish to simulate (in our case two black holes following
non-eccentric inspiral) {\em and} are in a suitable set of coordinates. 

The most widely used method to evolve black-hole-binary initial data is the 
moving-puncture method \cite{Campanelli:2005dd,Baker:2005vv}, which 
involves a modification of the Baumgarte-Shapiro-Shibata-Nakamura 
(BSSN)~\cite{Baumgarte:1998te,Shibata:1995we} formulation of the 3+1 
ADM-York Einstein equations~\cite{Arnowitt1962,York1979} combined with the 
``1+log'' ~\cite{Bona:1994dr} and ``$\tilde{\Gamma}$-driver'' gauge 
conditions~\cite{Alcubierre:2001vm,Alcubierre:2002kk}. 

As the name suggests, the data that are usually evolved with this method are puncture 
data~\cite{Brandt:1997tf}, whereby black holes are represented on the numerical 
grid by compactified wormholes. However, when these data are evolved
using the standard moving-puncture method the numerical
slices lose contact with the extra asymptotically flat wormhole ends, and quickly
asymptote to cylinders of finite areal radius located within the horizon
of each black hole. That the data evolve to these ``trumpets'' was 
realized in~\cite{Hannam:2006vv}, in which an analytic stationary
trumpet endstate was derived and shown to agree with numerical
results.

That work suggested a new form of initial data, based on trumpets. 
It was shown in~\cite{Hannam:2006xw} that maximally-sliced trumpet 
data can easily be
constructed numerically based on the solution first presented 
in~\cite{Estabrook:1973ue}, and that these data are indeed time independent
in a moving-puncture simulation. These data represent the first
non-trivial test solution for most current black-hole evolution codes. 
It was later shown in~\cite{Baumgarte:2007ht} that an implicit form
of the same solution could be constructed analytically, and 
in~\cite{Hannam:2008sg} the corresponding solution for the 1+log-sliced
case was found.

We have presented a detailed study of Schwarzschild wormholes and
trumpets in~\cite{Hannam:2008sg}, with a focus on constructing and
evolving Schwarzschild trumpet puncture data. This work extends
that study to boosted, spinning and binary trumpets. As we described 
in the concluding section of~\cite{Hannam:2008sg}, ideal binary 
puncture data will be in trumpet form, 1+log-sliced (or satisfy whatever
slicing condition is ultimately used to evolve them), and represent
true boosted Schwarzschild or Kerr black holes (i.e., will be free of 
the junk radiation that plagues all current binary simulations). As a first
step in a larger research programme to attempt to achieve that goal, 
we deal here with only the first point in our list of requirements: that the 
data be in trumpet form. The data we construct will not meet \emph{any}
of the other requirements: they will be maximally (not 1+log) sliced, 
and they will be conformally flat, meaning that they include essentially
the same junk radiation as standard puncture data. As such, this
work is a proof-of-principle exercise that demonstrates that is 
feasible to produce binary trumpet data. Along the way a number of 
new issues arise that are not present in the wormhole case, and variants
of these issues may recur in efforts to produce yet more general data. 

We will start with a brief summary of wormholes, trumpets and punctures in
Sec.~\ref{sec:background}, then discuss in some detail the maximal slicing
case in spherical symmetry in Sec.~\ref{sec:schwarzschild} as an
example for our analytical setup to construct trumpet data and for our
numerical algorithm.
We then extend the trumpet-puncture construction to boosted (Section~\ref{sec:boost})
and spinning (Section~\ref{sec:spin}) Bowen-York black holes, and provide an
outline of a proof for both existence and uniqueness of these solutions. In 
Section~\ref{sec:rad} we estimate the junk-radiation content of these initial-data
sets, before moving on to binary data in Section~\ref{sec:binary}. The ultimate
goal is of course to produce data that can be used in black-hole-binary 
simulations, and in Section~\ref{sec:evolution} we evolve a binary 
data set and compare with the corresponding standard wormhole-puncture
results. We close with a discussion on the next steps to producing 
optimal initial data for moving-puncture simulations.

\section{Background: a brief summary of wormholes, trumpets and punctures}\label{sec:background} 

\subsection{Wormhole puncture data}

Consider a constant-time slice of the Schwarzschild spacetime. Write the standard 
Schwarzschild solution in isotropic coordinates, i.e., \begin{equation}
 ds^2 = -\left( \frac{1 - \frac{M}{2r}}{1 + \frac{M}{2r}} \right)^2 dt^2 +
\psi^4 ( dr^2 + r^2 d \Omega^2  ),  \label{eqn:isotropicSchw} 
\end{equation} and the isotropic coordinate $r$ is related to the Schwarzschild
areal radial coordinate $R$ by
\begin{equation}
R = \psi^2 r, \label{eqn:isoR}
\end{equation} and $\psi = 1 + M/2r$ is a conformal
factor. Now the data $(\gamma_{ij},K_{ij})$
on any $t = constant$ slice are given by $\gamma_{ij} = \psi^4 \eta_{ij}$ 
(where $\eta_{ij}$ is the flat-space metric in the chosen coordinate system)
and $K_{ij} = 0$. The fact that the physical spatial metric can be related to the
flat-space metric using only the conformal factor indicates that the solution is
\emph{conformally flat}.

We see immediately from Eqn.~(\ref{eqn:isoR}) that the slice does not reach the 
physical singularity at $R=0$, or even penetrate the black-hole horizon at $R=2M$. 
In fact, the coordinate range $r\in[0,\infty]$ contains two copies of the Schwarzschild
spacetime exterior to $R=2M$: one copy in $r\in[0,M/2]$ and the other in $r\in[M/2,\infty]$.
These coordinates therefore represent the exterior Schwarzschild spacetime as
a wormhole, and this is most clear when viewed in an embedding diagram like
that shown in Fig.~1 of~\cite{Hannam:2008sg}. 

The advantage of these slices for numerical relativity is that the entire exterior
space can be represented on $R^3$ without any need to deal
explicitly with the physical singularity of the black hole, or to ``excise'' any 
region of the computational grid. The point $r=0$, which is commonly referred to as a
``puncture''~\cite{Brandt:1997tf}, represents a second copy of spatial infinity,
but the solution is well-behaved there, except for the conformal factor $\psi$,
which diverges as $1/r$. 

We can write initial data for multiple Schwarzschild black holes simply by 
modifying the conformal factor to $\psi = 1 + \sum_i m_i/(2r_i)$, where the
$m_i$ parametrize the mass of the $i$th black hole, and the $i$th puncture
is located at $r_i = 0$~\cite{Brill:1963yv}. Furthermore, one may imbue these 
black holes with 
linear and angular momentum by providing a non-zero extrinsic curvature. 
If we retain the property of conformal flatness and choose the extrinsic curvature 
to be trace-free ($K=0$, or \emph{maximal slicing}), then there exist solutions
of the momentum constraint for boosted and/or spinning black holes; these are
the Bowen-York solutions~\cite{Bowen:1980yu}. The solution is provided only
in the conformal space, and is related to the physical extrinsic curvature by 
\begin{equation}
K_{ij} = \psi^{-2} \tilde{A}_{ij},
\end{equation} where here $\tilde{A}_{ij}$ is the Bowen-York solution. 
Now, however, the conformal factor is not known analytically,
and can only be found by solving numerically the Hamiltonian constraint, 
\begin{equation}
\tilde{\nabla}^2 \psi + \frac{1}{8} \psi^{-7} \tilde{A}_{ij} \tilde{A}^{ij} = 0.
\label{eqn:HC}
\end{equation} The most convenient way to solve (\ref{eqn:HC}) is by the
``puncture method''~\cite{Brandt:1997tf}, which is to realize that the 
solution can be constructed conveniently in terms of a (typically small) 
correction $u$ to the Brill-Lindquist solution,
\begin{equation}
\psi = 1+ \sum_i \frac{m_i}{2r_i} + u.  \label{eqn:BLpsi}
\end{equation} Since the Brill-Lindquist conformal factor is in the kernel of 
the flat-space Laplacian, the Hamiltonian constraint is now an equation for the 
correction function $u$: \begin{equation}
\tilde{\nabla}^2 u + \frac{1}{8} \psi^{-7} \tilde{A}_{ij} \tilde{A}^{ij} = 0. \label{eqn:HCpunc}
\end{equation} Furthermore, the function $u$ is sufficiently regular over all of
$R^3$ that (\ref{eqn:HCpunc}) is in the form of a nonlinear elliptic equation that
is straightforward to solve by a number of standard methods. This approach is
used to construct the majority of black-hole-binary initial data used in current
numerical simulations, and the elliptic solve is performed either with mesh-refinement
finite-difference solvers~\cite{Brown:2004ma} or, in most cases, 
by an elegant single-domain spectral approach~\cite{Ansorg:2004ds}, which we
will adopt for the work presented here.

A key property of the  Bowen-York family of solutions to the momentum constraint is that
the values of the momentum and angular 
momentum of the spacetime (and thus in some sense 
the momenta and spins of the black holes) can be prescribed {\em before} solving for the
conformal factor.

Two further properties of these data deserve particular attention here, and we will 
return to them when comparing these wormhole puncture data with our new 
trumpet puncture data in the following sections. 

First, Bowen-York black holes
are conformally flat, which is not the case for 
either a true boosted Schwarzschild black hole, or a Kerr black hole, 
or a boosted Kerr black hole. Since it is usually a boosted Schwarzschild or
Kerr black hole that we really wish to describe, these data are often 
described as the desired physical objects plus some ``junk''; the junk
represents a valid part of a solution of Einstein's equations, but it is not
a part that we would expect to occur physically, and can be interpreted
as unphysical gravitational wave content. As the data 
evolve forward in time, the junk either falls into the black hole or 
radiates away, quickly leaving precisely the physical situation that was
intended in the first place, albeit with slightly different physical parameters. 

In practice (i.e., in black-hole-binary simulations) this junk radiation causes
two problems. One is that it introduces noise into the numerical simulation, which 
can affect the numerical accuracy. This point is clearly illustrated in~\cite{Boyle:2007ft}. 
The other is that it limits the physical black-hole spin that can be achieved.
When the spin angular momentum of the Bowen-York black hole is extremely 
high, most of the angular momentum manifests itself as junk, and after
that junk has either fallen into the black hole or radiated away, we are left
with a Kerr black hole that has spin no higher than $a/M = S/M^2 \leq 0.93$
\cite{Dain:2002ee,Dain:2008ck,Lovelace:2008tw}; we will confirm this with
high-precision numerical simulations, bounding the final Kerr parameter at 
$a/m \leq 0.929$.
This property of Bowen-York data preclude their use to study very highly 
spinning black holes, which may in fact be the most common 
astrophysically~\cite{Volonteri:2004cf,Gammie:2003qi,Shapiro05}, and 
we must turn to other types of data --- see, for example,
\cite{Hannam:2006zt} for the construction and evolution of spinning but non-boosted
puncture data, and~\cite{Lovelace:2008tw} for non-conformally-flat black-hole
initial data where the interior of the black hole is excised.

The other property of Bowen-York puncture data that we want to highlight
is the calculation of the black-hole mass. 
Having produced data for two black holes, we would
like to know what their masses are; although the parameters $m_i$ parametrize
the black-hole masses, the black-hole mass equals the mass parameter only in
the case of a single Schwarzschild black hole, i.e., the original Schwarzschild
solution in isotropic coordinates. 

In any other case, we typically estimate the black-hole mass by two methods. 
One is to calculate it from the area of the apparent horizon. This requires that
we first locate the apparent horizon, which can be computationally expensive 
(although fast and efficient solvers exist, for example~\cite{Thornburg:2003sf}).
The other method is to make an inversion transformation at each puncture
and calculate the ADM mass at that black hole's extra asymptotically flat end,
and to treat this quantity as the black-hole mass. For a binary system, this
mass estimate is given by \begin{equation}
M_i = m_i \left( 1 + u_{0,i} + \frac{m_i m_j}{2 D_{ij}} \right), \label{eqn:puncMass}
\end{equation} where $D_{ij}$ is the coordinate separation between the 
two punctures, and $u_{0,i}$ is the value of the correction function $u$ at 
the $i$th puncture. 
Remarkably, this expression is found to agree within numerical error with the 
mass calculated from the apparent horizon~\cite{Tichy:2003qi,Brugmann:2008zz},
although we will see in Section~\ref{sec:rad} that this can only be expected
to hold for boosted black holes, or black holes with small spins.

\subsection{Trumpet puncture data}

Bowen-York puncture data were first constructed long before stable
numerical simulations of black-hole binaries were possible, and 
were useful in both mathematical relativity~\cite{Beig:1993gt,Beig:1994rp,Dain:2001ry} 
and in studies of initial 
data~\cite{Brandt:1997tf,Baumgarte:2000mk,Hannam:2003tv,Hannam:2005ed,Hannam:2005rp}. 
However, with the advent of the moving-puncture 
method~\cite{Campanelli:2005dd,Baker:2005vv} it was found that wormholes
may not be the most suitable topology for black-hole initial data. 

In a moving-puncture simulation, the numerical slices quickly lose contact
with the extra asymptotically flat ends, and instead asymptote to 
cylinders of finite areal radius \cite{Hannam:2006vv,Hannam:2006xw,Brown:2007tb,Garfinkle:2007yt,Hannam:2008sg,Ohme:2009gn}, or ``trumpets''. This suggests that it 
would be more natural to construct initial data in trumpet form from the 
outset. 

To date this has only been done for a single Schwarzschild black 
hole. 
The question addressed in this paper is, How can we generalize
the wormhole puncture procedure to produce \emph{trumpet} punctures for
black-hole binaries? For a single maximally sliced Schwarzschild black hole, 
the trumpet data can be put in a form similar to the wormhole isotropic coordinates,
where now the conformal factor behaves as $\psi \sim \sqrt{3M/2r}$ near
the puncture. However, the full conformal factor is not known analytically
(except as an implicit equation in terms of the Schwarzschild radial
coordinate $R$)~\cite{Hannam:2006vv,Hannam:2008sg}. 
This means that it is not straightforward to 
superpose two trumpets as with the Brill-Lindquist solution in the 
wormhole case. And it is not obvious how the introduction of the 
Bowen-York extrinsic curvature (which, if we retain conformal flatness
and maximal slicing, remains a valid solution of the momentum
constraint), affects the behavior of the conformal factor near the 
puncture, or the physical properties of the data. Finally, without the
presence of extra asymptotically flat ends, we lose the simple procedure to 
estimate the black hole's mass from Eqn.~(\ref{eqn:puncMass}). These
are the issues that we address in this work.

In Section~\ref{sec:schwarzschild} we describe in more detail the 
maximal Schwarzschild trumpet, and use it to illustrate our more
general method for producing single-trumpet data.

\section{Maximal Schwarzschild trumpet}\label{sec:schwarzschild}
\label{sec:schwTrumpet}
\subsection{Constructing a conformal-factor ansatz for trumpet data}

The basis of this work are data that represent a maximal slice of the
Schwarzschild spacetime with a trumpet topology. The first hints of
this representation of Schwarzschild were given by Estabrook {\it
  et al.}~\cite{Estabrook:1973ue} in 1973, but it wasn't until the
development of the moving-puncture
method~\cite{Campanelli:2005dd,Baker:2005vv} in 2005, and a subsequent
understanding of the dynamical behavior of the numerical
slices~\cite{Hannam:2006vv} in that method, that it was realized that
the maximal Schwarzschild trumpet could be expressed in a simple
form~\cite{Hannam:2006xw}, and could in turn be written in the
``puncture'' isotropic coordinates suited to moving-puncture
simulations~\cite{Hannam:2006xw,Baumgarte:2007ht}.

For a single Schwarzschild black hole with mass $M$, the conformal
initial data in Cartesian coordinates are \begin{eqnarray*}
\tilde{\gamma}_{ij} & = & \delta_{ij}, \\
\tilde{A}_{ij}^S &  = & \frac{C}{r^3} \left( 3 n_i n_j  - \delta_{ij} \right), \\
K & = & 0, \\
\alpha & = &  \sqrt{1 - \frac{2M}{R} + \frac{C^2}{R^4}}, \\
\beta^i & = & \frac{ x^i \alpha C}{R^3},
\end{eqnarray*} where $C = \sqrt{27/16} M^2$, $R$ is the Schwarzschild
radial coordinate, $r = (x^2 + y^2 + z^2)^{1/2}$ is the isotropic
radial coordinate, and $n_i = x_i /r$ is the outward-pointing normal
vector. All that remains to fully specify the initial data is a valid
conformal factor $\psi$ that maps these data to the physical space,
i.e., \begin{eqnarray*}
\gamma_{ij} & = &\psi^4 \tilde{\gamma}_{ij} \\
K_{ij} & = & \psi^{-2} \tilde{A}_{ij}^S + \frac{1}{3} \psi^4
\tilde{\gamma}_{ij} K \\
R & = & \psi^2 r.
\end{eqnarray*} The conformal factor must satisfy the Hamiltonian
constraint and asymptote to $\psi \rightarrow 1$ as $r \rightarrow
\infty$. A numerical solution of the Hamiltonian constraint for these
data was first presented in~\cite{Hannam:2006xw}, and an analytic
solution (albeit an implicit solution in terms of $R$, not $r$) given
in~\cite{Baumgarte:2007ht}. 

To illustrate the method that we will 
use for more general cases, and to test our elliptic solver, we will again 
solve the Hamiltonian constraint numerically. Our boundary conditions are that 
$\psi \rightarrow 1$ as $r \rightarrow \infty$, and 
$\psi \sim \sqrt{3M/2r}$ as $r \rightarrow 0$; 
the latter condition ensures that we have a trumpet topology. 

In order to solve the Hamiltonian constraint, we start with an ansatz for 
$\psi$ that includes the required asymptotic behavior.
We write the 
full conformal factor that solves the Hamiltonian constraint as 
\begin{equation}
\psi = \psi_{s} + u,
\end{equation}
where $ \psi_{s}$ incorporates the desired asymptotics.
The Hamiltonian constraint for this problem is \begin{equation}
\tilde{\nabla}^2 u = - \frac{1}{8} \psi^{-7} \tilde{A}_{ij} \tilde{A}^{ij} - \tilde{\nabla}^2 \psi_{s},
\label{eqn:SingleHC}
\end{equation} where $\tilde{\nabla}^2$ represents the Laplacian with respect 
to the flat background metric, and it is understood that $\tilde{A}_{ij} = \tilde{A}_{ij}^S$, 
although this is the form of the Hamiltonian constraint that we will deal with
for all choices of $\tilde{A}_{ij}$ throughout this paper.

One easy way to incorporate the  asymptotic behavior is to apply weight
functions to the two asymptotic conditions, $$
\psi_{s}(r) = w_1(r) \sqrt{\frac{3M}{2r}} + w_2(r)
$$ such that \begin{eqnarray*}
w_1(0) = 1, \ \ \ \ w_1(\infty) = 0, \\
w_2(0) = 0, \ \ \ \ w_2(\infty) = 1. 
\end{eqnarray*} The weight functions we choose are \begin{eqnarray*}
w_1(r) & = & \frac{1}{1 + r^4}\,, \\
w_2(r) & = & \frac{r^4}{1 + r^4}.
\end{eqnarray*} These have the property that at each end of the slice the conformal factor's
lowest-order deviation from the required behavior is at fourth order.

Consider now the behavior of the conformal factor near the puncture. We assume the leading order terms to be of the form \begin{equation}
\psi =  \frac{A}{r^{1/2}} + B r^n. \label{eqn:psiAnsatz}
\end{equation} If we insert this
ansatz into the Hamiltonian constraint, we have \begin{eqnarray*}
\tilde{\nabla}^2 \psi & = & -  \psi^{-7} \frac{81 M^4}{64 r^6} \\
\Rightarrow - \frac{A}{4 r^{5/2}}  + B n (n+1) r^{n-2} & = & \frac{81 M^4}{64 A^7 r^{5/2}} \times \\
&  & \left( 1 - \frac{7B}{A} r^{n+1/2} + ... \right),
\end{eqnarray*} where we have expanded about $r=0$ on the right-hand side.
Equating  coefficients of $r$, we find that $A = \sqrt{3M/2}$ (as we 
expect). We also find that for a consistent solution $n = \sqrt{2} - 1/2 = 0.9142...$ 
and $B$ remains undetermined. We therefore see that divergent terms near
the puncture do exactly cancel, and the next-to-leading order term goes to
zero. However, this next-to-leading order term goes to zero with a non-rational
power of $r$ (which was also noted in~\cite{Bruegmann:2009gc}), and this may
limit the accuracy of a spectral solution to (\ref{eqn:SingleHC}). If this is the 
case, we may also include the $r^{\sqrt{2}-1/2}$ behavior into our ansatz. 

An implicit solution of $\psi$ in terms of the Schwarzschild radial coordinate $R$ 
is given in~\cite{Baumgarte:2007ht}, as is an implicit solution of 
$r(R)$. 
If we combine these as $(\psi(R) - \sqrt{3M/2r(R)})/r(R)^{\sqrt{2}-1/2}$,
and take the limit as $R \rightarrow 3M/2$, we can determine the coefficient $B$
in our ansatz above. We find that \begin{equation}
B = \left( \frac{3M}{2} \right)^{3/2} \left( M + \frac{3M}{2 \sqrt{2}} \right)^{-1-\sqrt{2}}.
\end{equation} If necessary, we may now use \begin{equation}
\psi_{s}(r) = w_1(r) \left( \frac{A}{\sqrt{r}} + B r^{\sqrt{2}-1/2} \right) + w_2(r),
\end{equation} as the ansatz in our numerical solution of the Hamiltonian constraint.

To summarize, we have two choices of conformal factor ansatz that we may adopt,
and which we denote by,
\begin{eqnarray}
\psi  =  \psi_{s1}(r) + u & = & w_1(r) \sqrt{\frac{3M}{2r}} + w_2(r) + u\,,   \label{eqn:psis1} \\
\psi  =  \psi_{s2}(r) + u & = & w_1(r) \left( \frac{A}{\sqrt{r}} + B r^{\sqrt{2}-1/2} \right) \nonumber\\
       & & + w_2(r) + u\,.  
\label{eqn:psis2}
\end{eqnarray}

\subsection{Numerical solution of the Hamiltonian constraint}

In order to solve the equations numerically, we have written a code to
solve systems of nonlinear elliptic equations with general finite difference
methods in three spatial dimensions. In this work we will only utilize this
solver with pseudospectral discretizations, representing the solution
by Fourier series in (periodic) angular coordinates, and as Chebyshev 
polynomials otherwise. The solver has been developed as a Mathematica package,
it uses the Mathematica {\tt LinearSolve} function with a Krylov method
and ILU preconditioner to solve Linear systems, and Newton iteration to
deal with nonlinearities. This approach has allowed us to develop a very
flexible spectral elliptic solver from scratch, in order to achieve good
performance even for the larger grids we use in this paper. We consistently
use sparse matrix objects and generate compiled code using Mathematica's 
{\tt CompiledFunction} for
certain key functions which operate on individual matrix elements.

The elliptic solver uses compactified coordinates
$(X,Y,\phi)$, with $X \in [-1,1]$, $Y \in (-1,1)$
and $\phi \in (-\pi,\pi)$. In all cases that involve a single black hole, we transform to 
these coordinates from spherical polar coordinates with $r = (1-X)/(1+X)$ and 
$Y = \cos(\theta)$, so that $X = -1$ corresponds to $r \rightarrow \infty$ and $X = 1$ 
corresponds to $r = 0$. In order for the coefficients of the Laplacian operator to be
sufficiently smooth over the entire domain, the entire equation is weighted by a factor
\begin{equation}
w_3(X,Y,\phi) = \frac{(1+X)^3 (1-Y^2)}{ (1-X)^2 }.
\end{equation}

\begin{figure}[t]
\centering
\includegraphics[width=80mm]{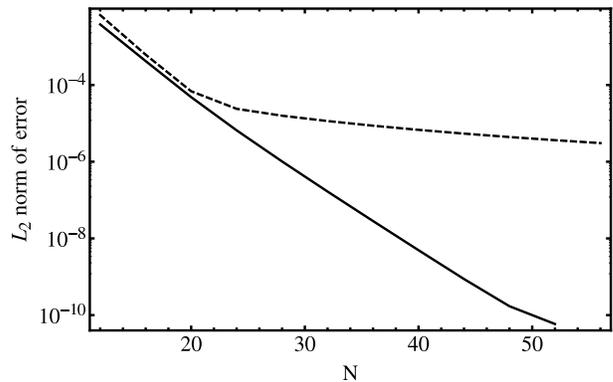}
\caption{The $L_2$ norm of the error in the solution function $u$ for a maximal
Schwarzschild trumpet. The dashed line shows the error when using the ansatz
(\ref{eqn:psis1}), 
while the solid line shows the error when using the ansatz (\ref{eqn:psis2}), 
which includes the next-to-leading order behavior in the conformal
factor near the puncture.
}
\label{fig:l2errorS}
\end{figure}

The accuracy of the numerical method is demonstrated in Fig.~\ref{fig:l2errorS},
which shows the $L_2$ norm of the error between the numerical and
analytic solutions as a function of the number of collocation points $N$. 
(The same number of points is chosen in each direction, although since
this solution is spherically symmetric, the solution varies only along the 
$X$ direction.) It is clear from Fig.~\ref{fig:l2errorS} that the spectral
convergence is lost for $N>20$ when the ansatz $\psi_{s1}$ is used,
but remains up to at least $N=48$ where the next-to-leading order
behavior is included in $\psi_{s2}$. 

\begin{figure}[t]
\centering
\includegraphics[width=80mm]{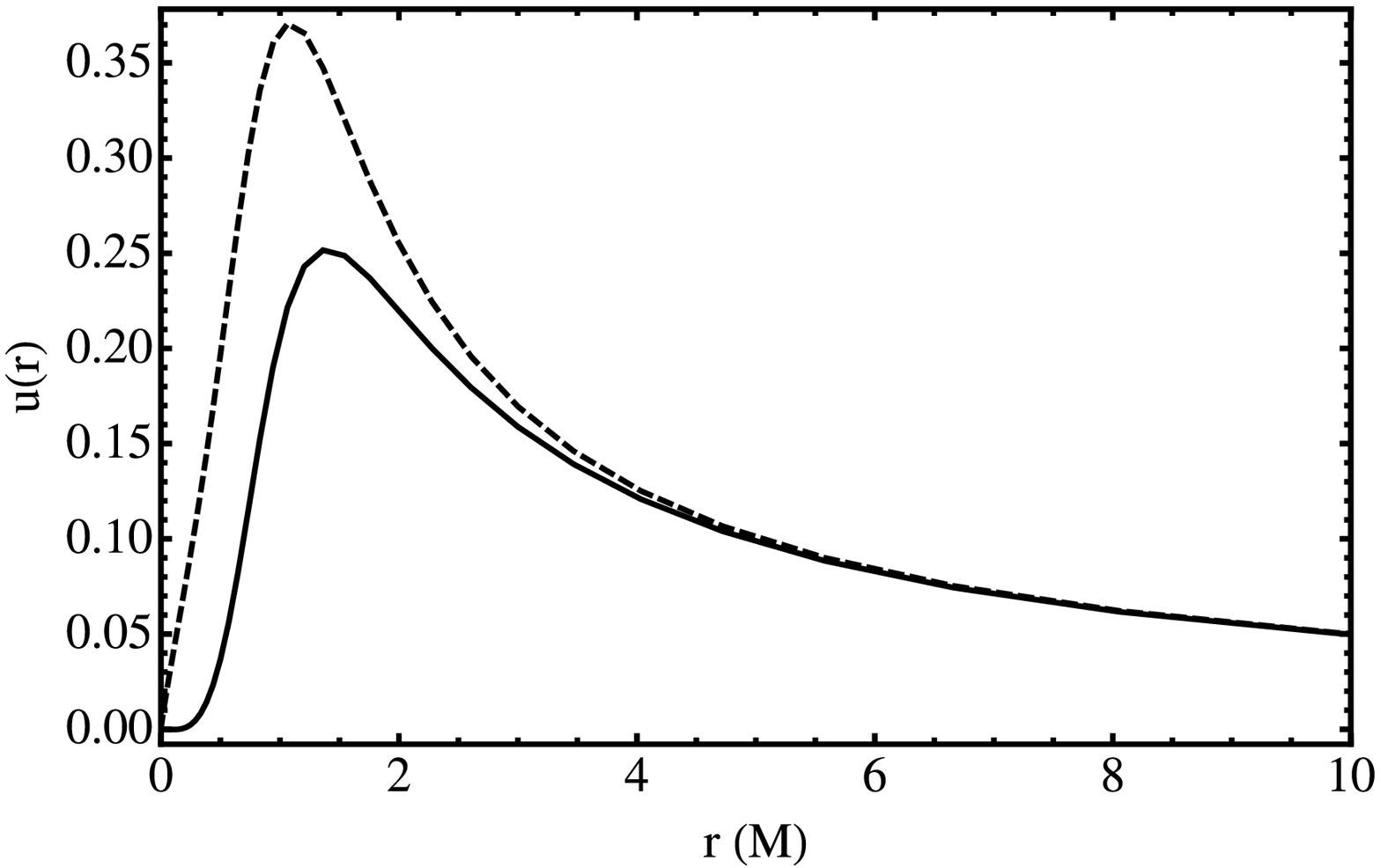}
\includegraphics[width=80mm]{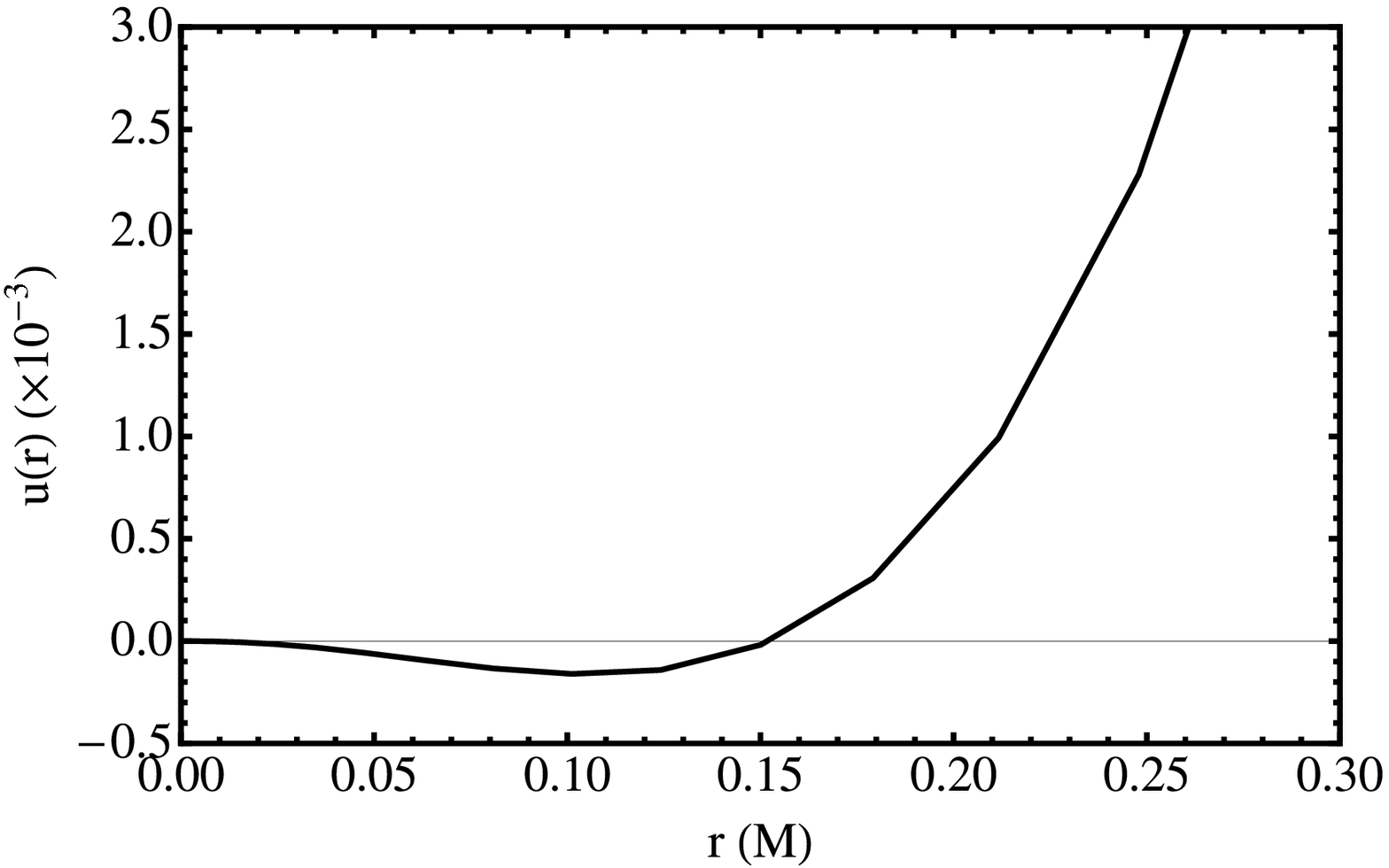}
\caption{
The correction function $u$ for the maximal Schwarzschild trumpet, for
the solution with $N=52$ collocation points. The solution using the $\psi_{s1}$
ansatz (\ref{eqn:psis1}) is shown with a dashed line, and the solution 
using the $\psi_{s2}$ ansatz (\ref{eqn:psis2}) is shown with a solid line. 
The second panel zooms into the region near the puncture,
to illustrate that the $\psi_{s2}$-based solution smoothly approaches zero there, 
and is well resolved; the $\psi_{s1}$ solution is poorly resolved in this 
region.
}
\label{fig:uSchwarzschild}
\end{figure}

The numerical solution $u$ is shown in Fig.~\ref{fig:uSchwarzschild}. Solutions
using both the $\psi_{s1}$ and $\psi_{s2}$ ans\"atze are shown. The
second panel in the figure zooms into the region near the puncture. In 
this figure the solution was produced using the ansatz with $\psi_{s2}$. 
We can see that the function smoothly
approaches zero at the puncture, and is well resolved by the numerical 
method. The $\psi_{s2}$-based solution is not well resolved near the puncture
and is not included in the second panel. The data in this plot are from solutions with 
$N=52$ collocation points.

\section{Single boosted Bowen-York trumpet} \label{sec:boost}

We now consider a single trumpet with linear momentum. 

To do this we add to the conformal extrinsic curvature the Bowen-York 
solution for a single
black hole with linear momentum $P^i$, \begin{equation}
\tilde{A}_{ij}^{BY}(r;\mathbf{P}) = \frac{3}{2r^2} \left( P_i n_j + P_j n_i - (\delta_{ij} - n_i n_j) P^k n_k \right),
\end{equation} so that the total conformal extrinsic curvature is \begin{equation} 
\tilde{A}_{ij}^{BYT} (r;\mathbf{P}) = \tilde{A}_{ij}^S(r) + \tilde{A}_{ij}^{BY}(r;\mathbf{P}).
\end{equation} The superscript ``BYT'' is a reminder that this is a Bowen-York trumpet. 

Note the asymptotic behavior of $\tilde{A}_{ij}^{BY}$ as $r \rightarrow 0$: it 
diverges as $1/r^2$. Since the trumpet extrinsic curvature diverges faster, as 
$1/r^3$, it dominates the Hamiltonian constraint near the puncture, and so
determines the behavior of the solution. In particular, this means that the 
trumpet form $\psi \sim \sqrt{3M/2r}$ remains.

Consider the general form of $\tilde{A}_{ij} \tilde{A}^{ij}$ for
the boosted case (in the following we will suppress the angular dependence
of the functions for simplicity): the contribution from the Schwarzschild trumpet 
diverges as $r^{-6}$, the Bowen-York contribution diverges as $r^{-4}$, 
and the cross terms diverge as $r^{-5}$, so we have \begin{equation}
\tilde{A}_{ij} \tilde{A}^{ij} = \frac{A_4}{r^4} + \frac{A_5}{r^5} + \frac{A_6}{r^6}.
\end{equation} 

Near the puncture, we can write the inverse conformal factor
term as \begin{eqnarray*}
\psi^{-7} & = & ( A r^{-1/2} + u )^{-7} \\
& = & A^{-7} r^{7/2} ( 1 + u \, r^{1/2}/A )^{-7} \\
& = & A^{-7} r^{7/2} - 7 A^{-8} u \, r^4 + O(r^{9/2}),
\end{eqnarray*} where $A$ is the same quantity that was introduced in 
Eqn.~(\ref{eqn:psiAnsatz}). We can now write out the source term of the Hamiltonian 
constraint as \begin{equation}
\frac{1}{8} \psi^{-7} \tilde{A}_{ij} \tilde{A}^{ij} = \sum_{i=0}^2 \frac{D_i}{r^{1/2+i}}
+ \sum_{i=0}^2 \frac{u \, D'_i}{r^i}. \label{eqn:sourceExp}
\end{equation} The $D_2$ term is the one that diverges as $r^{-5/2}$ and
is canceled by a corresponding term from the Laplacian of $\psi_{s}$, as 
described in Section~\ref{sec:schwTrumpet}. The remaining terms all 
result in contributions to $u$ with positive powers of $r$, and which therefore
go to zero at the puncture, \emph{except} for the $D'_2$ term, which can
in principle lead to a contribution that diverges as $\ln r$. We note that
such a term also appears in the Schwarzschild case (with our choice of
ansatz), but there we know that $u=0$ at the puncture, and so none of
the $D'_i$ terms contribute to the solution. Fortunately, we will see in the 
existence proof that we present below that the same is true in the boosted
case.  In the coordinates of 
our elliptic solver, the puncture $r=0$ is located on the entire coordinate
plane $X=-1$, and so there we can simply impose that either $u=0$
or $u'=0$, and thus prevent the solver from producing unphysical divergent
terms. 

Before proceeding, we will show that solutions to this problem exist 
and are unique. Note that while construction of a numerical solution gives
evidence for the existence of a solution to the continuum equations, uniqueness
is not easy to verify numerically, and an analytical proof is highly desirable.
While the uniqueness proof is general, the existence proof requires a more
detailed analysis of the Hamiltonian constraint, and in the spinning case
we will deal with only a single trumpet (we do however expect that the same procedure
can be generalized to multiple spinning and boosted black holes).

We first prove uniqueness. 
Assume we have two positive solutions, $\psi_1$ and $\psi_2$. 
Subtract the equations to get 
$$
\tilde\nabla^2(\psi_1 - \psi_2) = -{1 \over 8}\tilde{A}_{ij}\tilde{A}^{ij}(\psi_1^{-7} - \psi_2^{-7}).
$$
We assume $\psi_{12} = \psi_1 - \psi_2$ goes to zero at both ends (they satisfy the same boundary conditions, and we saw in the preceeding discuss that there are no other
divergent terms in the solution). If $\psi_{12}$ is not identically zero, it must have a positive maximum or a negative minimum. Neither of these is compatible 
with the equation (leading to different signs on the left and right hand side).

We now provide the outline of an existence proof. 
The maximum principle tells us that a solution, if it exists,
cannot have an interior minimum.
As $r \rightarrow \infty$ our boundary condition is that $\psi \rightarrow 1$, 
and so a solution, if it exists, satisfies $\psi \ge 1$. 
Therefore
$$\tilde \nabla^2\bar\psi = - {1 \over 8} K_{ij}K^{ij}$$
is a supersolution, i.e., it satisfies $\tilde{\nabla}^2(\bar\psi - \psi) \le 0$ and $\bar{\bar\psi} = 1$ 
is a subsolution, i.e., $\tilde \nabla^2(\bar{\bar\psi} - \psi) \geq 0$, and, of course $\bar\psi > 0$. Finally, the solution with linear momentum $P = 0$ lies between. Therefore, as we change $P$ the solution is trapped between the sub- and supersolution. The supersolution diverges as $r^{-4}$ 
as $r \rightarrow 0$, proving that the true solution cannot have any divergence stronger
than $r^{-4}$, and in particular that there are no logarithmic divergences. This allows us
to posit an ansatz for $\psi$ consistent with the allowed blow-up powers, and then 
check by consistency with the full Hamiltonian constraint which of those survive to the
full solution; and this leads to the $r^{-1/2}$ behavior determined in 
Section~\ref{sec:schwTrumpet}. This completes our outline of an existence proof,
which holds for single and multiple-black-hole solutions. The only complication 
arises when the trumpet has angular momentum, but we will deal with this case in
Section~\ref{sec:spin}.  
A more rigorous proof along the lines of that for the wormhole-puncture 
case \cite{Beig:1993gt,Dain:2001ry} remains to be constructed, and would be an interesting topic for 
future work. 

Having proved that solutions to this system exist and are unique, we now
must find them numerically. One potential problem that is apparent from 
Eqn.~(\ref{eqn:sourceExp}) is that the $D_i$ source terms involve half-integer
powers of $r$ near the puncture, which affects the 
accuracy of the elliptic solver. Concretely, the $D_0$ term will lead to
a $r^{3/2}$ contribution to the solution, which we expect to limit
the solver to 1.5-order accuracy near the puncture, 
and the $D_1$ term will lead to a $r^{1/2}$ contribution, which we
expect will limit
the solver the 0.5-order accuracy near the puncture \cite{Boyd00}, 
and appears at a 
lower order than the $r^{\sqrt{2}-1/2} \approx r$ term that we have already
accounted for in the $\psi_{s2}$ ansatz. 

\begin{figure}[t]
\centering
\includegraphics[width=80mm]{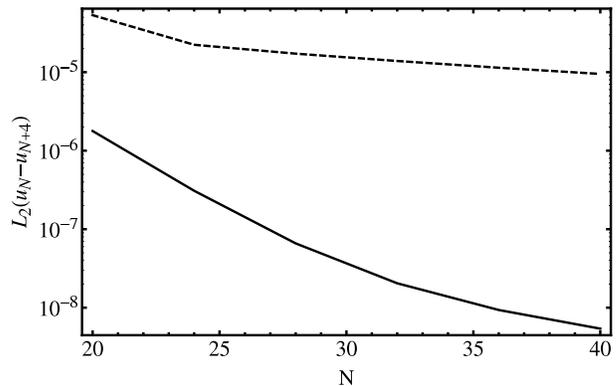}
\caption{
The error behavior of the Hamiltonian-constraint solution for a 
single boosted trumpet. The dashed line shows the convergence of the 
$L_2$ norm over the entire domain, while the solid line shows the $L_2$
norm for the region of the domain outside the black-hole horizon.
See text for more details. 
}
\label{fig:BoostError}
\end{figure}

These expectations are borne out in our results. Fig.~\ref{fig:BoostError} 
shows the 
convergence behavior of the $L_2$ norm for the entire solution. 
We find that the convergence is at less than first-order, consistent with the
half-order convergence predicted above. (Since we no longer have an 
analytic solution to compare with, we evaluate the convergence by comparisons
between solutions with successive numbers of collocation points. We chose to
sample $N$ in multiples of four, and therefore display the $L_2$ norm of 
$(u_{N+4} - u_N)$ in the figure.) However, if we
include in the $L_2$ norm only that part of the computational domain
that is outside the apparent horizon of the black hole (located approximately
at $r = 0.77m$), then the errors show exponential convergence
up to about $N = 32$. For higher numbers of collocation points the convergence
rate deteriorates, and for the larger values of $N$ shown in the figure the 
results are consistent with fourth-order convergence. This demonstrates
that the behavior near the puncture limits the accuracy of the solution, 
but that this limitation is essentially localized within the black hole.  

\begin{figure}[t]
\centering
\includegraphics[width=80mm]{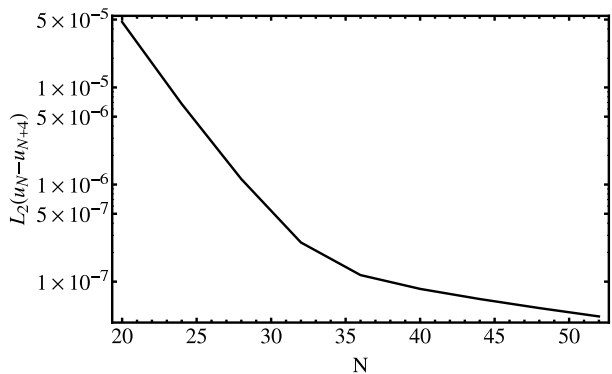}
\caption{
The error behavior of the Hamiltonian-constraint solution for a 
single boosted trumpet, with the $r^{-5}$ term removed from the source
term. The solution displays clean exponential convergence up to 
about $N = 32$, and then the convergence deteriorates to 1.5-order
 (see text).
}
\label{fig:BoostErrorX}
\end{figure}

The $D_1$ term is due to the $A_5/r^5$ term in $\tilde{A}_{ij} \tilde{A}^{ij}$
(which is in turn due to the cross-term between the Schwarzschild and Bowen-York
extrinsic curvatures). If we remove these cross terms from the source
function, we obtain the convergence behavior shown in Fig.~\ref{fig:BoostErrorX}; 
we now see, as expected, that for $N>32$ the convergence approaches
1.5-order over the entire domain, consistent with the earlier discussion.
Unfortunately, this solution does not represent the correct conformal
factor for a boosted Bowen-York trumpet puncture!

Although the inclusion of the Bowen-York extrinsic curvature limits the 
accuracy of our solver near the puncture, the solution is still very accurate 
over most of the computational domain, and is anyway accurate enough
for most practical purposes everywhere. If one wished to produce yet
more accurate solutions, one option would be to use a coordinate
transformation from $r$ to $X$ that lead to the solution near the 
puncture being expanded in powers of $r^{1/2}$. However, for the 
purposes of this paper, such accuracy is not required, and we simply 
make this observation for future use.

\section{Single spinning Bowen-York trumpets} 
\label{sec:spin}

The construction of a solution for a single spinning Bowen-York puncture
trumpet is complicated by the fact that the Bowen-York extrinsic curvature
for a spinning black hole diverges as $1/r^3$ near the puncture. In this 
case, the behavior of the conformal factor near the puncture \emph{will}
be modified by the presence of the Bowen-York term. However, we 
will show that it is possible to determine the angular dependence of the 
divergent term in the conformal factor from a \emph{local} one-dimensional
ODE, which can be easily solved to construct the appropriate ansatz for
a full numerical solution.

\subsection{Angular dependence at the puncture} 

For convenience in what follows, we will express the problem in 
spherical coordinates. We will assume that the conformal factor
now behaves as $\psi \sim D(\theta)/ \sqrt{r}$ as $r \rightarrow 0$. The
square of the conformal extrinsic curvature that appears in the
Hamiltonian constraint is now \begin{equation}
A^2 \equiv \tilde{A}^{ij} \tilde{A}_{ij} = \frac{6C^2}{r^6} + \frac{18 S^2(1 - \cos^2 \theta)}{r^6},
\label{eqn:AsqrBY}
\end{equation} where $S$ is the angular momentum of the black hole. 
Note that in this case there are no cross terms. 

To extend our earlier existence proof to the spinning case, we need to take into 
account the change in the divergent term in the conformal factor.
We start by proving a monotonicity condition for the Hamiltonian constraint 
with these boundary conditions. More precisely, Let $A_1^2 = A^2(C,S_1)$ and 
$A_2^2 = A^2(C,S_2)$, where $S_2 > S_1$.  Since $A_2^2 \ge A_1^2$, then 
$\psi_2 \ge \psi_1$. This means that, if we fix $C$ and pump up 
$S$, the conformal factor monotonically increases.

The proof is as follows. Subtract the two solutions to 
get
$$ \tilde{\nabla}^2(\psi_2 - \psi_1) + \frac{1}{8} [A_2^2 \psi_2^{-7} -   A_1^2 
\psi_1^{-7}] = 0. $$
Now multiply across by $r^m$ where $m$ lies between $1/2$ and $1$, 
and find an equation for $\xi = r^m( \psi_2 - \psi_1)$.
We find that \beq
\tilde{\nabla}^2 \xi - \frac{m}{r} \partial_r \xi -  \frac{m - m^2}{r^2} \xi 
 + \frac{r^m}{8} [A_2^2 \psi_2^{-7} -   A_1^2 \psi_1^{-7}] = 0.  \label{eqn:xi}
\eeq
We can see that $\xi$ vanishes both at $r = 0$ and at infinity. 
The quantity $\xi$ 
can never be negative because, if it were, it would have a negative 
minimum, and this cannot happen. Let us assume that it does have such a 
negative minimum. Let us see what happens to Eqn.~(\ref{eqn:xi}) 
at that point. We have $\nabla^2 \xi \ge 0, - m/r \partial_r \xi = 0, - (m - m^2)/r^2 
\xi > 0,$ and $+ r^m/8 [A_2^2 \psi_2^{-7} -   A_1^2 \psi_1^{-7}] \ge 
0$. The last term is the only slightly tricky term. If $\xi < 0$, then 
$\psi_2 < \psi_1$ and $\psi_2^{-7} > \psi_1^{-7}$. Since we assume 
$A_2^2 \ge A_1^2$, this term is also non-negative and the sum cannot 
add up to zero.

Now we want to consider how $D(\theta)$ behaves, where we assume 
$\psi = D(\theta)/\sqrt{r} + O(\sqrt{r})$ near the origin.
When we substitute into the Hamiltonian constraint, 
we get the following equation for $D(\theta)$:
\begin{equation}
D'' + \frac{D'}{\tan(\theta)} - \frac{1}{4}D
+ \frac{1}{8 D^7} \left[6C^2 + 18S^2(1 - \cos^2 \theta) \right] 
= 0,          \label{eqn:ode}
\end{equation}
where $D''$ is second derivative with respect to $\theta$. 
This is defined on the interval $0 \le \theta \le \pi$, but will be symmetric 
around $\pi/2$. At a maximum we have
$$
D^8 < 1/2[6C^2 + 18S^2(1 - \cos^2 \theta)],
$$
while at a minimum we have
$$
D^8 > 1/2[6C^2 + 18S^2(1 - \cos^2 \theta)].
$$
Therefore the maximum should occur at $\pi/2$ and the minimum at $\theta 
= 0$ and $D$ satisfies
$$
3C^2 \le D^8 \le  [3C^2 + 9S^2].
$$ These upper and lower bounds allow our earlier existence proof to 
go through unchanged. 

Eqn.~(\ref{eqn:ode}) should be read as a one-dimensional 
second-order equation for $D(\theta)$ on the interval 
$0 \le \theta \le \pi/2$, with Neumann boundary 
conditions, i.e., $D' = 0$ at both ends.

A solution of Eqn.~(\ref{eqn:ode}) provides the necessary information to 
construct a single spinning Bowen-York puncture trumpet. The most important
feature of Eqn.~(\ref{eqn:ode}) is that it is \emph{local}: we need only solve
a simple one-dimensional ODE in order to calculate the requisite boundary 
information --- 
\emph{regardless of the linear momentum of the black hole, and regardless of the
presence or otherwise of other black holes in the data}.

\subsection{Solution of the nonlinear angular-dependence ODE} \label{sec:ode}

\begin{figure}[t]
\centering
\includegraphics[width=80mm]{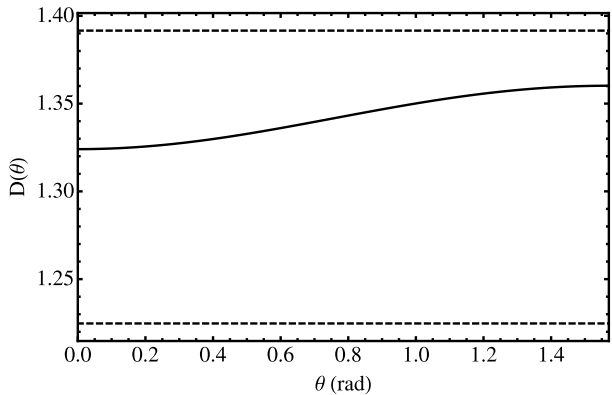}
\caption{The solution $D(\theta)$ for $C^2 = 27/16$ and $S = 1.0$. The
  upper and lower bounds, $D_{lower}^8 = 3C^2$ and $D_{upper}^8 = 3C^2
+ 8S^2$ are shown with dashed lines.}
\label{fig:DJ1}
\end{figure}

We solve the nonlinear ODE Eqn.~(\ref{eqn:ode}) by linearizing and
solving iteratively. The average of the upper and lower bounds is used as
an initial guess. A simple application of the {\tt NDSolve} function 
in Mathematica suffices to produce an accurate solution. The solution
for $S/M^2=1$ is shown in Fig.~\ref{fig:DJ1}; the function $D(\theta)$ is
seen to lie well within the upper and lower bounds derived in the 
previous section.

\begin{figure}[t]
\centering
\includegraphics[width=80mm]{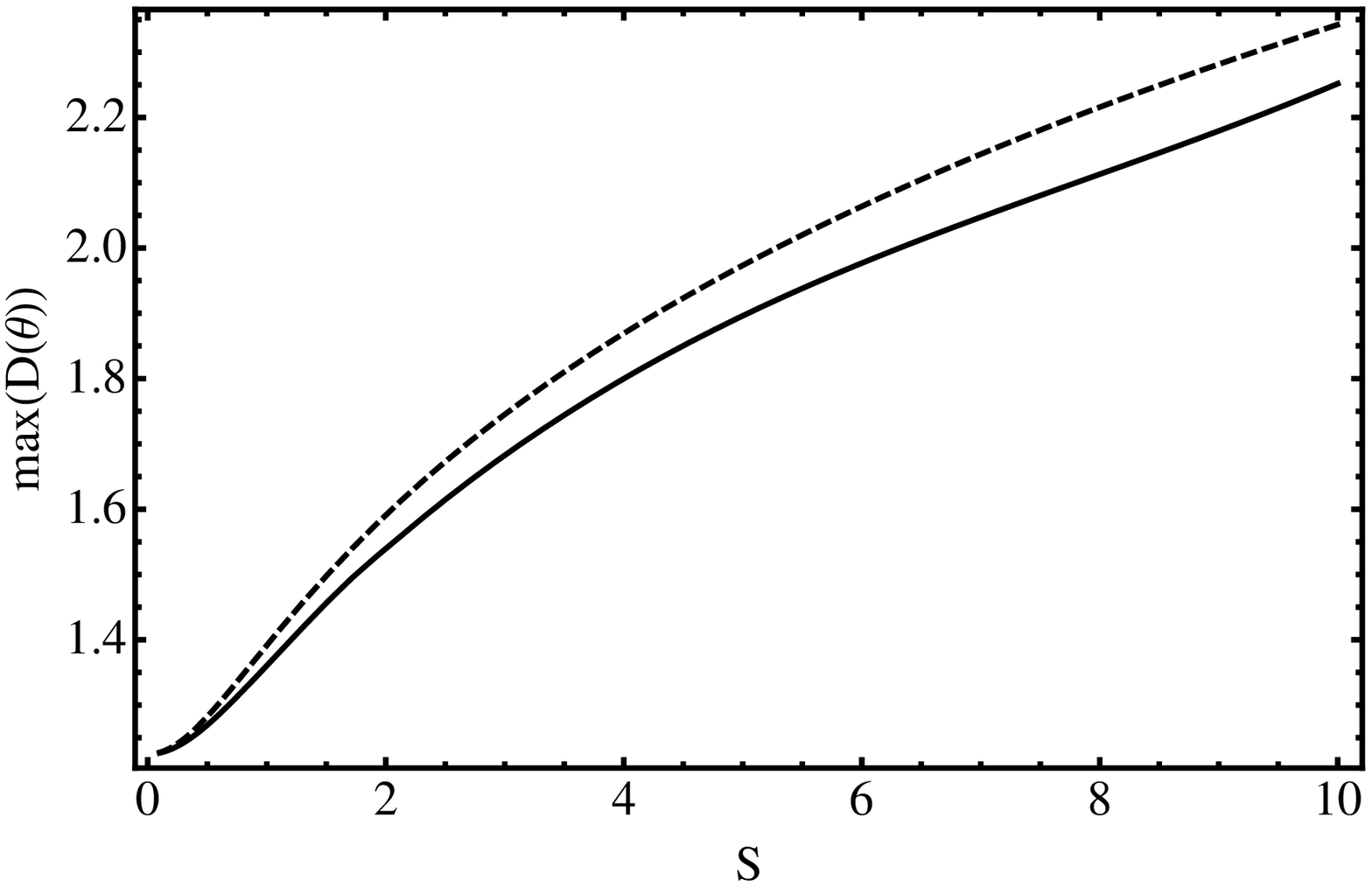}
\includegraphics[width=80mm]{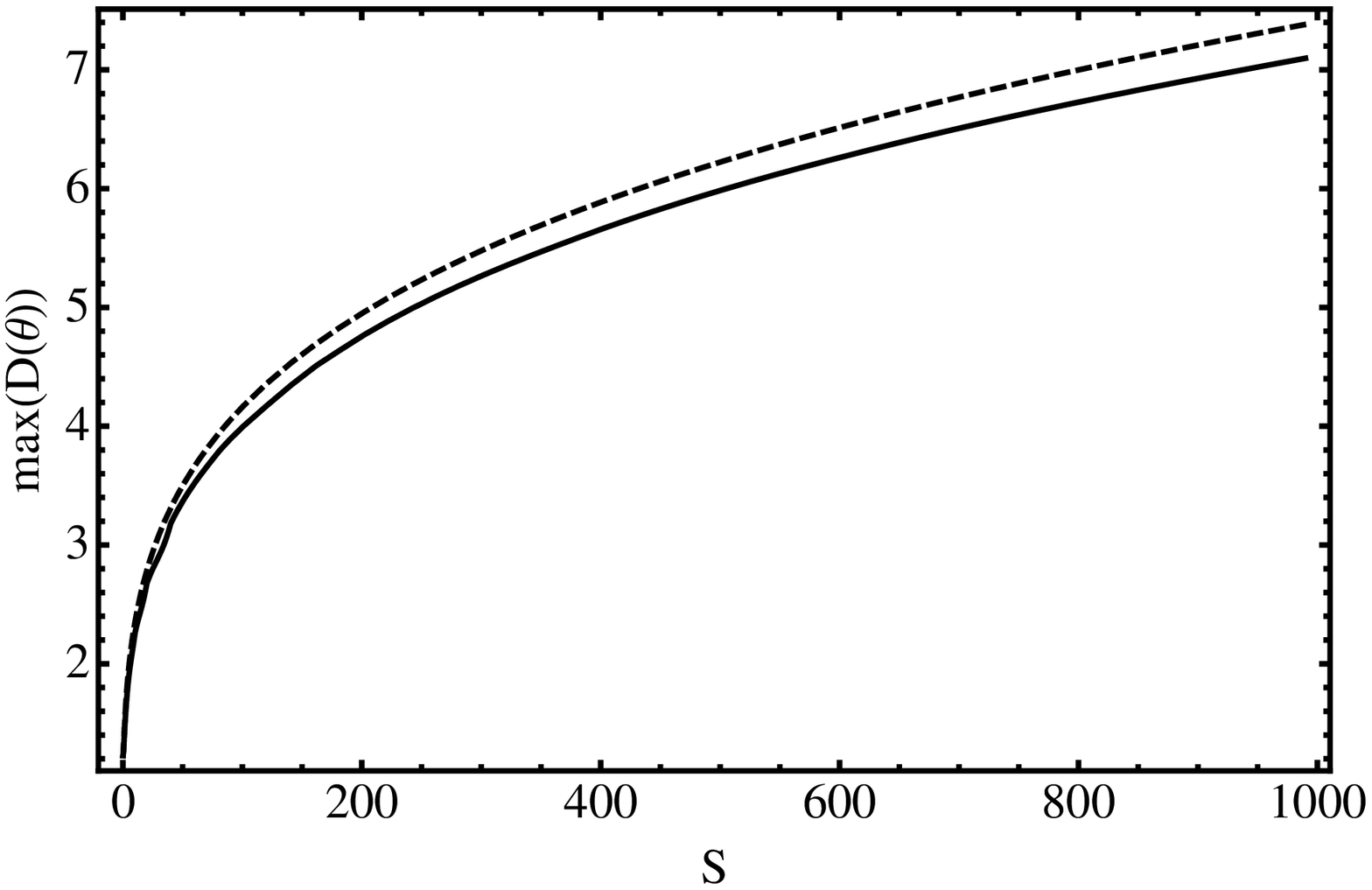}
\caption{Values of the maximum value of $D(\theta)$ (at $\theta =
  \pi/2$) as a function of the angular momentum $S$, shown with a
  solid line. Also shown as a
  dashed line is the upper bound. The maximum behaves as $S^{1/4}$ for large
  $S$.}
\label{fig:Dmax}
\end{figure}

Figure~\ref{fig:Dmax} shows the maximum value of $D(\theta)$, which
occurs at $\theta = \pi/2$, as a function of the angular momentum
$S$. The figure shows the upper bound on the solution, $(3C^2 +
9S^2)^{1/8}$, for comparison. The maximum behaves as
expected, i.e., grows as $S^{1/4}$ for large $S$. When $S$ is small,
the $3C^2$ term dominates, and the value approaches the Schwarzschild
value of $\sqrt{3/2}$. 

Now that we have calculated $D(\theta)$, we are able to solve the 
Hamiltonian constraint for both boosted and spinning Bowen-York
trumpets. The conformal-factor ansatz is now provided by replacing the 
$\sqrt{3M/2r}$ term in (\ref{eqn:psis1}) with $D(\theta)/\sqrt{r}$.
In the numerical procedure to solve the Hamiltonian constraint, 
the derivatives of $D(\theta)$ required
in the construction of $\tilde{\nabla}^2 \psi_s$ are trivial to calculate in our
{\tt Mathematica}-based solver, because $D(\theta)$ is available from the 
solution to (\ref{eqn:ode}) as an {\tt InterpolatingFunction} to whatever 
precision is required.

Note, however, that for the spinning case we do not know
the next-to-leading order behavior of the solution to the Hamiltonian constraint
(the coefficient  of the $r^{\sqrt{2}-1/2}$ term) as we did in the boosted case, and this
will restrict the accuracy of our solver to that given by the $\psi_{s1}$
ansatz in Section~\ref{sec:boost}, and of course the magnitude of 
this term will grow with the value of the angular momentum. For 
this reason, high accuracy is difficult to achieve for extremely high
values of the spin. For the data sets studied in this paper we
consider angular momenta no higher than $S = 10M^2$, 
which corresponds to $S/M^2 \approx 0.924$. We will now discuss 
the junk radiation content of our data sets in more detail.

\section{Radiation content of trumpet-puncture data}
\label{sec:rad} 

Bowen-York black holes can be considered as Kerr or boosted Schwarzschild 
black holes, plus some unphysical radiation content, which either falls into the black
hole or radiates away as junk radiation. We can estimate the radiation
content of the data as~\cite{CookPhD,Cook:1989fb} 
\begin{equation}
E_{rad} = \sqrt{E_{ADM}^2 - P^2} - M.
\end{equation} 

To evaluate this quantity we first need an estimate of the black hole's mass
$M$. The standard way to calculate this is via the area of the apparent horizon
of the black hole. We calculate the irreducible mass, $M_{irr} = \sqrt{A/16\pi}$ and 
then use the Christodoulou formula~\cite{Christodoulou:1970wf} to estimate 
the total mass of a black hole with angular momentum $S$,
\begin{equation}
M^2 = M_{irr}^2 + \frac{S^2}{4 M_{irr}^2}. \label{eqn:ahmass}
\end{equation} 

For boosted wormhole data, the black-hole mass can also
be estimated by calculating the ADM mass at the extra asymptotically flat end;
one can see by performing an inversion transformation on the Bowen-York
extrinsic curvature that its contribution at the extra end falls off as $r^{-4}$,
and therefore we expect that it contributes very little junk radiation in the 
second copy of the exterior space. This suggests that the ADM mass
evaluated at the second asymptotically flat end (i.e., at the puncture) will
provide a good measure of the mass, and this has been confirmed by 
numerical observations~\cite{Tichy:2003qi}, and the ``ADM puncture mass''
has become a standard tool in wormhole puncture 
data~\cite{Brandt:1997tf,Hannam:2005ed,Hannam:2005rp,Brugmann:2008zz}.

There are two drawbacks of the ADM puncture mass. One is that it does not 
provide a good estimate of the mass for spinning black holes, since in that case 
the Bowen-York extrinsic curvature has the same fall-off behavior at both 
asymptotically flat ends, $O(r^{-3})$, and contributes roughly the same junk radiation into 
both exterior regions. We have verified this in numerical tests, where
we find that the ADM puncture mass for spinning-Bowen-York-puncture
data sets equals the ADM mass calculated at spatial infinity to within
the numerical accuracy of the solver ($\approx 10^{-8}$).

The other disadvantage of the ADM puncture mass, which applies in
general to wormhole puncture data,
is that the mass cannot be prescribed {\it a priori}, because the relationship
between the mass parameter $m$ and the black-hole mass $M$ is nonlinear. 
In order to construct Bowen-York wormhole punctures with specific masses,
an iteration procedure must be used. 

The situation appears to be quite different in the trumpet case. Here the
mass parameter $m$ \emph{does} seem to prescribe the mass of the black 
hole, at least for boosted black holes. This is presumably related to the fact
that the Bowen-York extrinsic curvature does not affect the geometry of the 
trumpet, irrespective of the value of the linear momentum. This interesting 
(and useful) property of the boosted Bowen-York trumpet deserves further study.

The same cannot be the case for spinning
black holes, however, where the coefficient of the singular term in the
conformal factor is an angular function of the spin. We could propose a
mass based on the area of the trumpet, but this is not necessarily useful, 
because we do not know the relationship between the trumpet area and
the black-hole mass for spinning black holes. For spinning black holes
we must make use of the mass calculated from the area of the apparent 
horizon, Eqn.~(\ref{eqn:ahmass}).

We are now in a position to estimate the junk radiation content of our
boosted and spinning trumpet data sets. 

\begin{figure}[t]
\centering
\includegraphics[width=80mm]{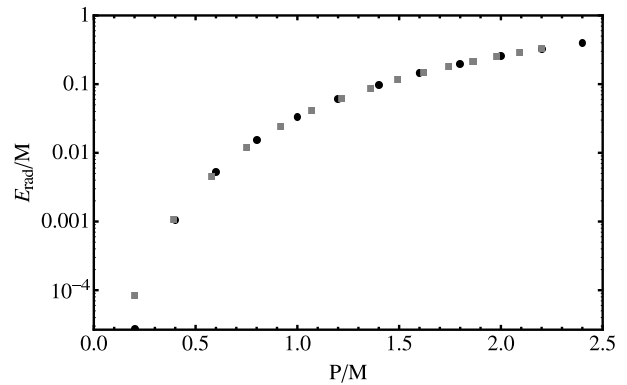}
\caption{
Estimate of the radiation energy content of boosted black-hole initial-data
sets. The grey squares indicate wormhole data, and the black circles indicate
trumpet data. The results for both Bowen-York trumpets and wormholes are shown. 
The results are identical at the level of accuracy of the data: as one might 
expect, the use of a trumpet versus a wormhole topology does not affect
the radiation content of the data. 
}
\label{fig:RadP}
\end{figure}

Fig.~\ref{fig:RadP} shows the estimate of the radiation content for 
boosted wormhole and trumpet initial-data sets. We see that the results
are almost identical for both classes of initial data. This also provides 
further evidence of the equivalence of the mass estimates that were 
used for each class of data. These results can further be compared with
those for other families of boosted Bowen-York 
data~\cite{CookPhD,Cook:1989fb,Hannam:2005ed}, for which the values of 
the junk radiation content appear to be very similar.

\begin{figure}[t]
\centering
\includegraphics[width=80mm]{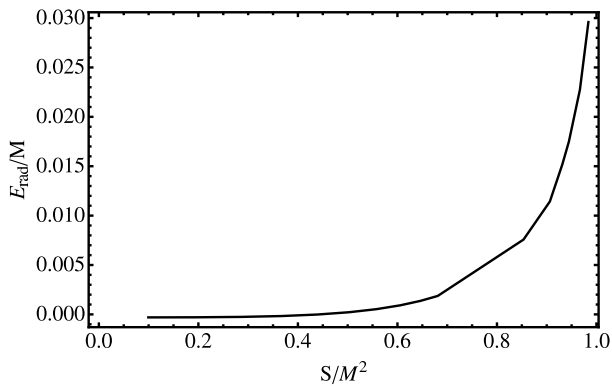}
\caption{
Estimate of the radiation energy content of spinning black-hole-trumpet 
initial-data sets, including the extreme limit, at which $S/M^2 = 0.9837$ and
$E_{rad} = 0.0296$, i.e., the junk radiation never consists of more than 2.96\% 
of the energy of the spacetime.
}
\label{fig:RadJ}
\end{figure}

Fig.~\ref{fig:RadJ} shows the same quantity estimated for spinning trumpet 
data sets. 
If we compare with the results in~\cite{CookPhD,Cook:1989fb} we see  
that the use of the trumpet
topology does not noticeably change the junk radiation content. 

It was pointed out in~\cite{Dain:2008ck,Dain:2008yu} that taking the limit 
as $m\rightarrow0$ while keeping $S$ fixed is equivalent to keeping $m$ fixed
and taking the limit $S\rightarrow \infty$. In other words, by simply removing the 
Schwarzschild trumpet term from the extrinsic curvature, we can construct 
data equivalent to the $S\rightarrow \infty$ limit. Furthermore, since we know that
the horizon is located at the puncture for these data, 
we can directly calculate the apparent-horizon
area to high accuracy from our angular function $D(\theta)$: 
\begin{eqnarray}
A & = & \lim_{r \rightarrow 0} \int \psi^4 r^2 \sin(\theta) d\theta d\phi \\
& = & 2 \pi \int D^4(\theta) \sin(\theta) d\theta.
\end{eqnarray}
We do this and find that  $S/M^2 = 0.9837$, in precise agreement with the 
results in \cite{Lovelace:2008tw}, although we note that via Eqn.~(\ref{eqn:ode}) one
can calculate this value to arbitrary accuracy. We also find that 
$S/M_{ADM}^2 = 0.928$, again in agreement with the results 
in \cite{Lovelace:2008tw}. These numbers provide
upper and lower bounds on the spin of the final Kerr black hole, after the junk radiation 
has left the spacetime. We evolved these data, and found that less than $0.05$\,\% of
the energy in the initial slice was radiated away, and therefore the rest of the junk
radiation falls into the black hole (in agreement with the observations 
in~\cite{Dain:2008ck}), and the final Kerr black hole has a spin 
parameter of $0.928 \leq a/m \leq 0.929$. Note also that
it follows from the results in \cite{Dain:2008ck,Dain:2008yu} that the 
high-angular-momentum limits of the wormhole and trumpet Bowen-York data
are equivalent. 

\section{Binary trumpets}
\label{sec:binary}

We now wish to construct data for two Bowen-York trumpets. The
linearity of the momentum constraint with $K = 0$ allows us to superimpose
any number of solutions: for each black hole we simply include both the 
Schwarzschild trumpet extrinsic curvature and the Bowen-York 
extrinsic curvature to obtain a valid solution of the momentum constraint. 
For black holes located at $\mathbf{r}_1$ and $\mathbf{r}_2$, the extrinsic curvature 
is therefore \begin{eqnarray}
\tilde{A}_{ij} & = & \tilde{A}_{ij}^S (\mathbf{r} - \mathbf{r}_1) +  \tilde{A}^{BY}_{ij} (\mathbf{r} - \mathbf{r}_1;\mathbf{P}_1) \nonumber \\
& & + \tilde{A}_{ij}^S (\mathbf{r} - \mathbf{r}_2) +  \tilde{A}^{BY}_{ij} (\mathbf{r} - \mathbf{r}_2;
\mathbf{P}_2).
\end{eqnarray} 

We once again need a suitable ansatz for the conformal factor. The first obvious choice is
to generalize the ansatz used for a single black hole and try
\begin{eqnarray}
\psi_s^{\rm guess} & = & w_1(r_1) \left[ \sqrt{\frac{R_{01}}{r_1}} +  R_{01}^{3/2} \left( m_1 + \frac{R_{01}}{\sqrt{2}} \right)^p r_1^q \right] \nonumber \\
&  & w_1(r_2) \left[ \sqrt{\frac{R_{02}}{r_2}} + R_{02}^{3/2} \left( m_2 + \frac{R_{02}}{\sqrt{2}} \right)^p r_2^q\right]  \nonumber \\
&  & + w_2(r_1) w_2(r_2), \label{eqn:binaryansatz}
\end{eqnarray} where $R_{0i} = 3m_i/2$, $p= -1-\sqrt{2}$ and $q = \sqrt{2}-1/2$,
and where the $w_2$ weightings are multiplied so that the resulting function is 
zero at each puncture, and asymptotes to unity far from the source. 

We saw in Sec.~\ref{sec:boost} that the requirement that $u=0$ at the puncture 
removed any logarithmically divergent terms from the solution, 
but this was possible only 
because the problematic part of the source term was linear in $u$: setting
$u=0$ removed that term. In the binary case, with the ansatz we have chosen, 
this is not necessarily so simple. Near one puncture (let us choose $r_1$) the
conformal factor behaves as $\psi = \sqrt{3m/2r} + F + u$, where $F$ is
the contribution from the second term in Eqn.~(\ref{eqn:binaryansatz}). This could
also generate a logarithmic term. One solution would be to determine
the appropriate value of $u$ at the puncture such that this term no longer
contributes (i.e., $u(r_1=0) = -F$), and enforce this in the solver, or hope 
that the solver finds that value \footnote{Research performed concurrently 
with that in this paper found that indeed the solver does appear to locate
this value~\cite{Immerman:2009ns}}. An
alternative solution is to choose an additional weighting factor so that 
in fact $A=0$, and to again impose our standard $u=0$ or $u'=0$
boundary condition at the puncture; this is the approach that we will follow.

To solve the Hamiltonian constraint numerically for binary trumpets, we 
adopt similar coordinates in our pseudospectral solver as developed
in~\cite{Ansorg:2004ds} for use with wormhole puncture data. For an
equal-mass binary with punctures located on the $x$-axis at $x = \pm b$,
we make the coordinate transformation \begin{eqnarray}
x & = & \frac{2 b \left(5 + X(2+X)\right) Y}{(1+Y^2)(3+X)(X-1)}\,,  \label{eqn:bipx} \\
y & = & \frac{4 b (1+X) (Y^2-1)\cos \phi}{(1+Y^2)(X^2+2X-3)}\,, \\
z & = & \frac{4 b (1+X) (Y^2-1) \sin \phi}{(1+Y^2)(X^2+2X-3)}\,. \label{eqn:bipz}
\end{eqnarray} In these coordinates $X = 1$ corresponds to spatial
infinity. The points $(X,Y) = (-1,\pm1)$ correspond to the puncture 
locations at $x = \pm b$. The line along the $x$-axis between the two
punctures is mapped to the plane $X=-1$. For a full description of this
coordinate system and its properties, the reader is referred 
to~\cite{Ansorg:2004ds}.

These coordinates make it particularly simple to apply additional weighting 
factors that remove at each puncture the 
contribution to the conformal factor ansatz from the other puncture. The weights
we choose  are $w_1 \rightarrow w_1 \cos\left[(\pi/4) (1\pm Y)\right]^4$.  

As an example, we construct data for the same configuration as in the 
``D10'' case studied in~\cite{Hannam:2007ik}: the punctures are located
at $x=\pm5M$, and the momenta are 
$\mathbf{P} = (\mp 9.80376 \times 10^{-4}, \pm 0.0961073 , 0)$. The
specific momenta are not important for this test; we simply choose the
same numbers to allow a direct comparison of the initial-data sets.

\begin{figure}[t]
\centering
\includegraphics[width=80mm]{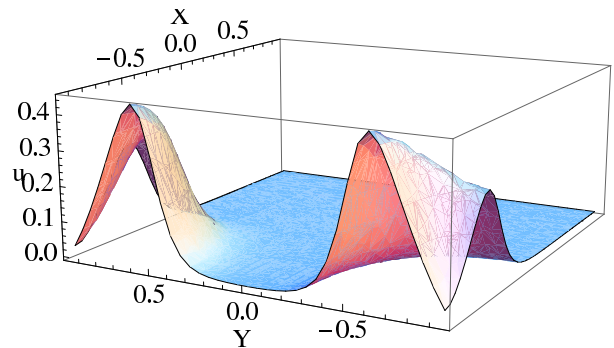}
\caption{
The function $u$ for the binary configuration described in the text,
represented in the bi-polar coordinates used in the pseudospectral
solver. Note that $X=1$ corresponds to spatial infinity, while $X=-1$,
$Y=\pm1$ are the puncture locations. 
}
\label{fig:BinaryU}
\end{figure}

The solution $u$ for this system is shown in Fig.~\ref{fig:BinaryU}, represented
in the coordinates (\ref{eqn:bipx}) -- (\ref{eqn:bipz}), along the plane 
$z=0$ ($\phi = 0$). 

\begin{figure}[t]
\centering
\includegraphics[width=80mm]{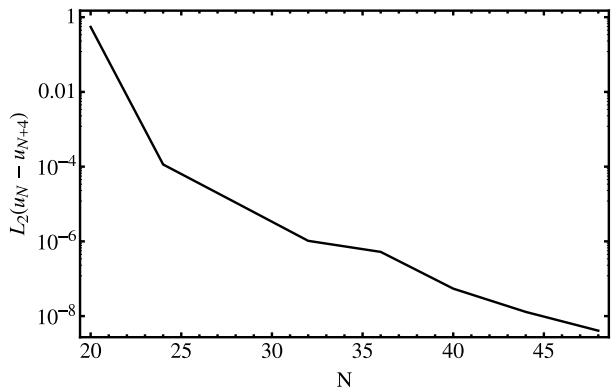}
\caption{
Error behavior for a binary configuration. The plot shows the $L_2$ norm
of the difference between solutions produced with $N$ and $N+4$ collocation
points. Only values along the $z=0$ plane are included in the calculation,
but since the punctures lie in this plane, this plot shows the dominant
error behavior for the solution. 
}
\label{fig:BinaryError}
\end{figure}

The convergence of the solver for these data is shown in Fig.~\ref{fig:BinaryError}.
The results indicate surprisingly good convergence in comparison to the
single-black-hole cases. This may be due to a cancellation in some other 
problematic terms in the binary case. For example, far from the binary
the Hamiltonian constraint source term will 
closely resemble that of a single spinning black hole; similar cancellation affects 
may play a role throughout the computational domain.

We can calculate the ADM mass of the system by noting that near
spatial infinity, $\psi \sim 1 + M_{ADM} / (2r)$, and obtain the ADM mass
from the radial derivative of $u$ as $r \rightarrow \infty$.  
As an indication of the accuracy of 
our solver, and of the level of difference between wormhole and puncture 
binary data, the ADM masses for the wormhole and puncture data with the 
same choice of black-hole mass, separation and linear momenta, were
0.9897136 and 0.989706, respectively. If we calculate the binding energies
($E_b = M_{ADM} - M_1 - M_2$) for these two data sets, they are 
$E_{b,{\rm wormhole}} = -0.0102864$ and 
$E_{b,{\rm trumpet}} = -0.0102939$. This demonstrates that these data
sets are physically extremely close --- with the added advantage in the 
trumpet case that the black-hole masses could be specified directly through
the mass parameter, while in the wormhole case they had to be
calculated by a nonlinear iteration 
procedure~\cite{Brugmann:2008zz,Hannam:2007ik}.

\section{Numerical evolution of the data}
\label{sec:evolution}

Having proposed and produced a new class of black-hole initial data, 
and claimed certain gauge and physical properties for them, we now need to
evolve a set of trumpet binary data and put our claims to the test. 
In particular, there are two questions we wish to answer: \begin{enumerate}
\item We expect that the trumpet data are in coordinates closer to those
preferred by the moving-puncture method than wormhole data; is this true? 
\item Do the wormhole and trumpet data describe the same 
\emph{physical} situation, or, in practical terms, do they produce the 
same gravitational-wave signal?
\end{enumerate}

We evolve the data using the same version of the BAM 
code~\cite{Brugmann:2008zz,Husa:2007hp} used to produce the 
results in~\cite{Hannam:2007ik}, with which we compare the 
gravitational waveform. In the notation of those works. we use the
same $N=64$ grid layout as used for the ``D10'' simulation; see 
Table~1 in~\cite{Hannam:2007ik}.

\subsection{Gauge changes}

The first question that we have posed above is difficult to answer. 
The data that we have produced 
are maximally sliced, while in the moving-puncture method one usually 
deals with 
1+log slicing, and the data will quickly cease to be maximally sliced and will
asymptote to their appropriate 1+log
form. In addition, the punctures are initially stationary, but will
pick up speed once the evolution begins; this constitutes yet another 
change of gauge. These gauge changes may be ``larger'' than those 
induced by the transition of wormhole data to trumpet form --- whatever
``larger'' means in the context of gauge changes. 

However, we can perform one simple test to quantify the change in gauge
between the two sets of data. In wormhole data, the apparent horizons of 
the two black holes are located on surfaces with coordinate radii close to 
$r \approx m/2$, where $m$ is the mass parameter in the wormhole puncture
conformal ansatz (\ref{eqn:BLpsi}). For trumpet data, on the other hand,
the horizon is at about $r \approx 0.78M$. If we evolve both wormhole and
trumpet data with a variant of 1+log slicing that will asymptote to maximal
slicing for a stationary spacetime, then we expect that the horizon radius will 
stay roughly
fixed in the trumpet case, while in the wormhole case it will increase quickly
to a value close to $r \approx 0.78M$. (The rapid expansion of the horizon 
early in simulations is standard in moving-puncture simulations; see, for example, \cite{Campanelli:2005dd,Baker:2005vv,Brugmann:2008zz}.)

\begin{figure}[tb]
\centering
\includegraphics[width=8cm]{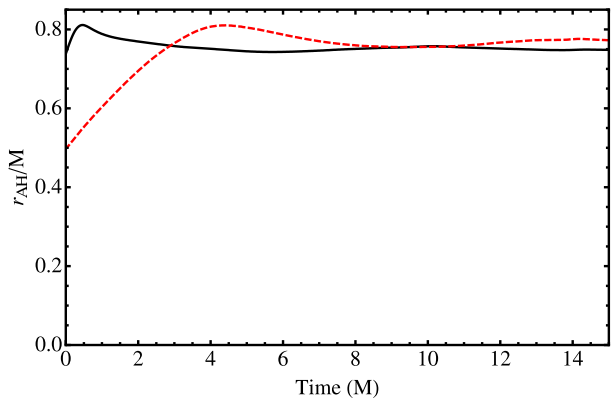}
\caption{Coordinate radius of the apparent horizon as a function of
  time, for one of the black holes in a binary evolution. 
  The data are initially maximally sliced. At early times the slicing will 
  deviate from $K=0$, but in a stationary situation would return to 
  maximal slicing within $t \approx 10M$. 
  As expected, the apparent-horizon radius shows much less
deviation for trumpet data (solid line) than for wormhole data
(dashed line). }
\label{fig:AHmotion}
\end{figure}

The slicing condition that approaches maximal slicing for a stationary 
solution is \begin{equation}
\partial_t \alpha = - 2 \alpha K,
\end{equation} i.e., the standard 1+log slicing used in
moving-puncture simulations, but {\it without} the shift term on the
left-hand side. With this gauge condition the data will deviate from 
maximal slicing at early times, but will again be approximately 
maximally sliced after about $t=10M$ of evolution~\cite{Hannam:2008sg}.
In addition we set $\eta = 0$ in the 
$\tilde{\Gamma}$-driver shift condition, to minimize additional gauge-related
growth in the horizon~\cite{Brugmann:2008zz,Hannam:2008sg}. 
The results are shown in Fig.~\ref{fig:AHmotion}, and are as expected:
in the wormhole case the horizon radius grows to about $0.75M$ within
$10M$ of evolution, while in the trumpet case the horizon radius 
remains close to that value at all times. The additional oscillations
may be due to other gauge effects, but are of much smaller magnitude
than the main effect we have just described. 

\subsection{Junk radiation}

We construct two sets of binary initial data (wormhole and trumpet)
for a binary with initial coordinate separation of $D=10M$. We adjust
the initial momenta such that both sets of data exhibit quasi-circular
inspiral. (We find that different values of the initial momenta are required
for each class of data; the reasons for these small differences are
at least partially due to the coordinate change made manifest by the different
apparent horizon sizes mentioned previously, and deserve
further investigation in future work.) 
We then evolve using standard
moving-puncture gauge choices, i.e., the full 1+log slicing condition,
$(\partial_t - \beta^i \partial_i) \alpha = -2 \alpha K$, and with
$\eta/M = 2$ in the $\tilde{\Gamma}$-driver condition.
We now wish to evaluate the 
differences in the gravitational-wave signal between simulations using
each data set.

\begin{figure}[t]
\centering
\includegraphics[width=80mm]{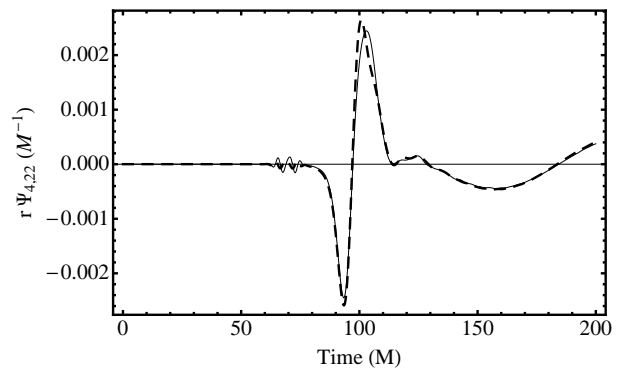}
\caption{
Junk radiation pulse from binary wormhole and trumpet data, 
with initial separation of $D = 10M$, 
and radiation-extraction radius of $R_{ex} = 90M$. 
The trumpet-data results
are shown with the thick dashed line, and the wormhole-data results are shown with 
the grey continuous line. 
As suggested by Fig.~\ref{fig:RadP}, the junk-radiation
content of both data sets is of comparable magnitude.
}
\label{fig:junk}
\end{figure}

The first point of comparison is the burst of junk radiation at the 
beginning of the simulation. Based on the results in Section~\ref{sec:boost},
we would expect that the junk radiation is the same in wormhole 
and trumpet data. Fig.~\ref{fig:junk} shows the pulse of junk radiation
in the spin-weight -2, $(\ell=2,m=2)$ mode of $r\Psi_4$, as calculated $R_{ex}=90M$
from the source. (Full details of the wave-extraction procedure used in the 
code are given in~\cite{Brugmann:2008zz}.) Although the junk pulses from the
two data sets are not identical, they are very similar; it is certainly not possible to 
definitively claim that one type of data contains less junk radiation than 
the other. 

We emphasize that this result is not merely a demonstration of a result that
we know to be true. The estimate of the radiation content of the initial-data
sets, based on the initial data alone, is no more than that: an estimate. It is
only by evolving the data in a full general-relativistic simulation that we can 
be certain that this (or any other) property that we claim for a new initial-data
set actually holds.

\subsection{Inspiral-merger-ringdown signal} 

We now consider the full inspiral-merger-ringdown GW signal generated
by the inspiral and coalescence of the two black holes. In this simulation
the binary completes about five orbits before merger.

\begin{figure}[t]
\centering
\includegraphics[width=80mm]{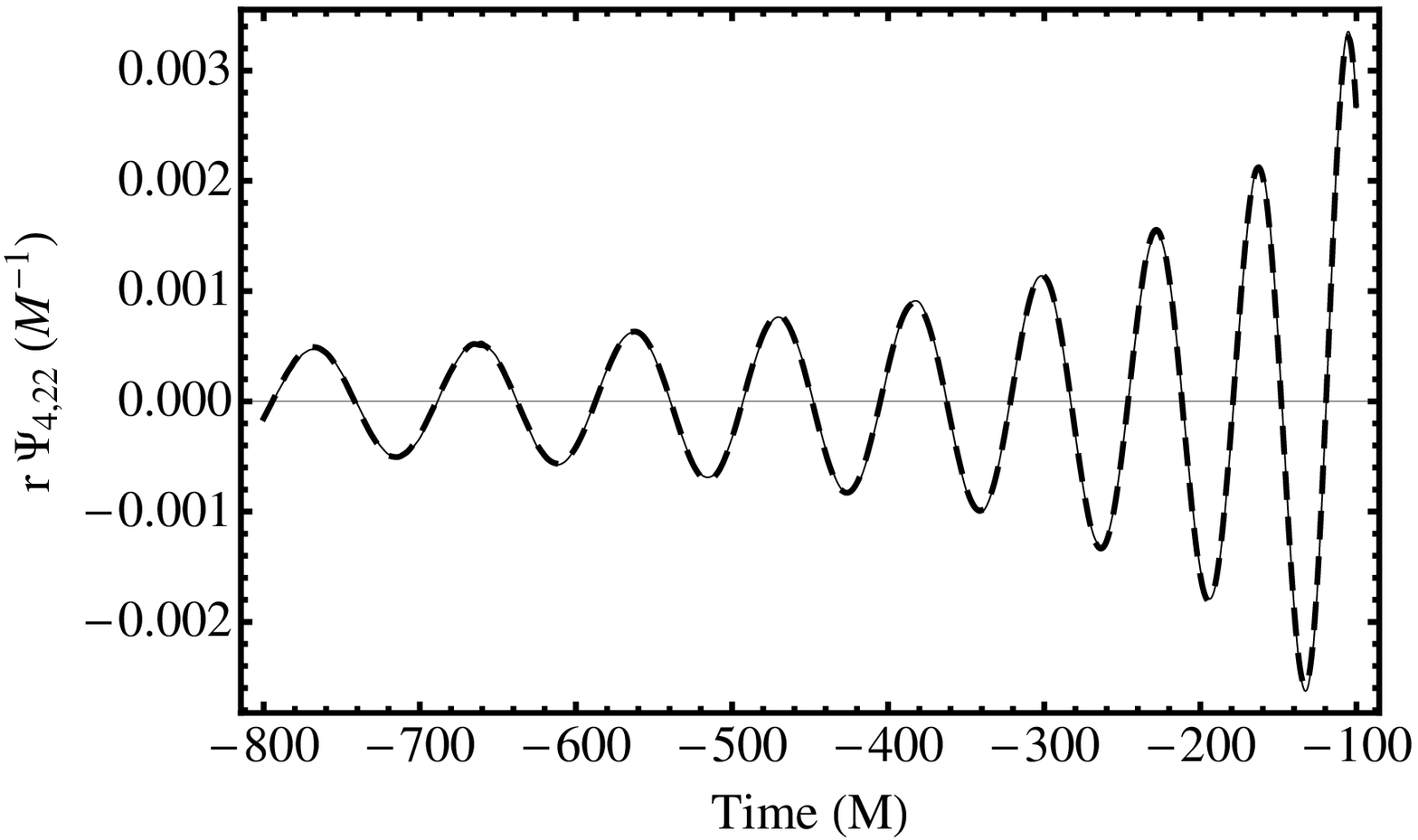}
\includegraphics[width=80mm]{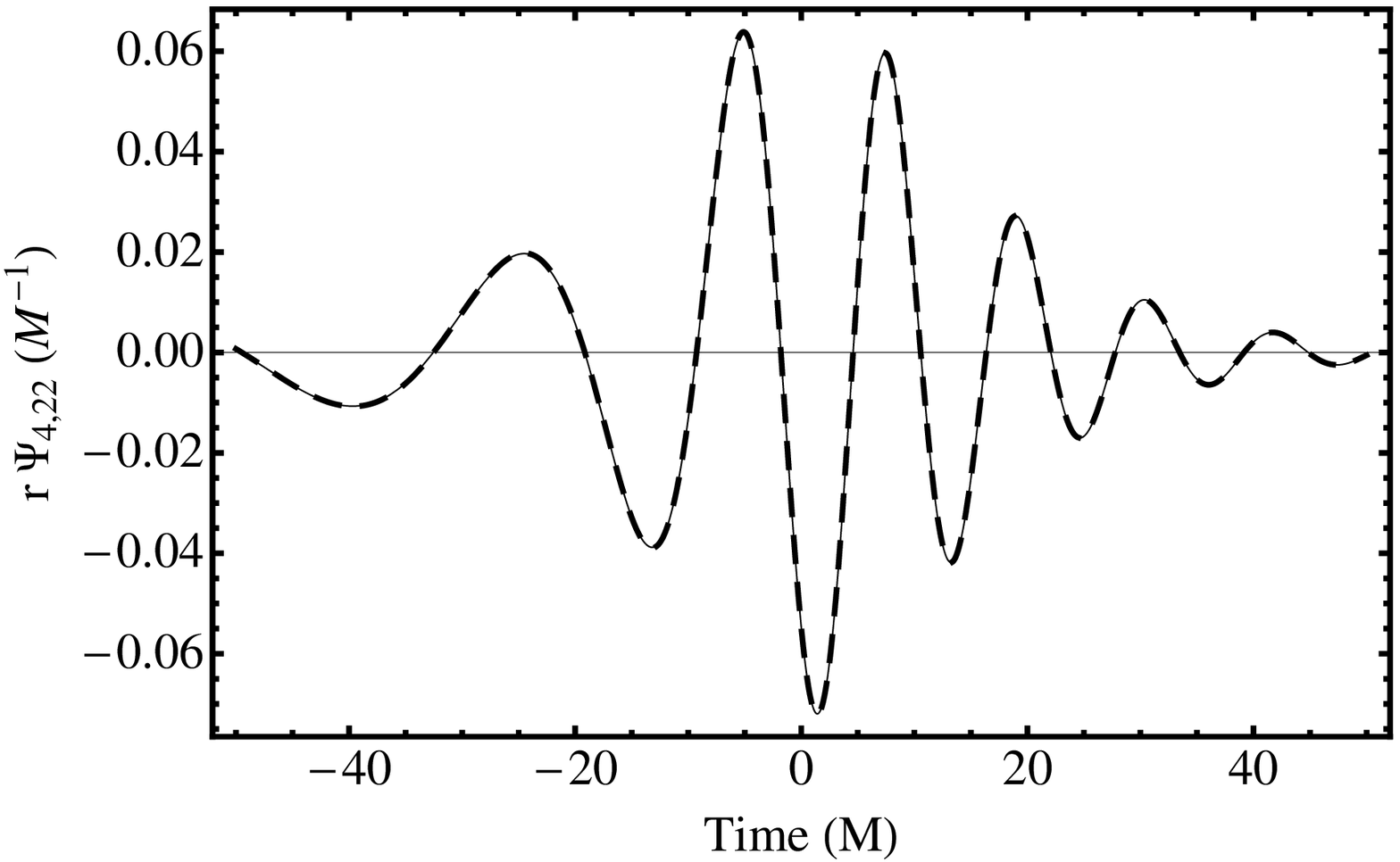}
\caption{Comparison of the inspiral and merger waveforms. The trumpet-data results
are shown with the thick dashed line, and the wormhole-data results are shown with 
the grey continuous line. A time and phase
shift have been applied so that the amplitude maxima occur at the same time,
at which time the waveforms are in phase }
\label{fig:psi4comp}
\end{figure}

We focus of the
dominant ($\ell=2,m=2$) spin-weighted spherical harmonic mode of
$r \Psi_4$, as extracted at $R_{ex} = 90M$ from the source. 
Fig.~\ref{fig:psi4comp} shows separately the inspiral and
merger-ringdown portions of the real part of $r\Psi_{4,22}$. 
(The plot begins after the 
junk radiation has passed through the $R_{ex}=90M$ radiation
extraction sphere.) The time has been shifted so that the maximum
amplitude occurs at $t=0$. 

The figure includes both the wormhole- and trumpet-data results. 
The results are indistinguishable, except for a very small
amount of de-phasing early in the signal, due to the slightly different
effective choice of initial parameters. 

\begin{figure}[t]
\centering
\includegraphics[width=80mm]{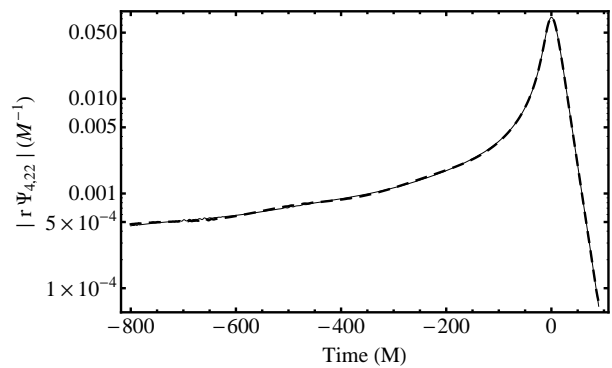}
\caption{The amplitude of $r\Psi_{4,22}$, as calculated from moving-puncture simulations 
of wormhole and trumpet puncture initial data, with initial separation of $D=10M$. 
The waveforms are shifted such that the maxima in the amplitude occur at the same time.
The thick dashed line shows the trumpet-data results, while the continuous grey line
shows the wormhole-data results.}
\label{fig:psi4amp}
\end{figure}

Fig.~\ref{fig:psi4amp} shows the amplitude of $r\Psi_{4,22}$ for the wormhole
and trumpet data. In this case the lines can be distinguished due to the slightly 
different eccentricities present in the two data sets. Once again it is clear,
however, that the two waveforms agree extremely well; they certainly agree
well within the error levels discussed in the recent Samurai project~\cite{Hannam:2009hh},
which demonstrated that waveforms that agree to this level are well within 
the accuracy requirements for detection and parameter estimation with first-
and second-generation ground-based GW  detectors. 

\section{Discussion}

In this work we have extended the puncture method to produce 
\emph{trumpet} data for boosted, spinning and binary black holes
based on the Bowen-York extrinsic curvature.
In the boosted case the generalization is straightforward, and in
the spinning case a simple one-dimensional nonlinear ordinary 
differential equation must be solved to determine the angular 
dependence of the asymptotic trumpet geometry. 

We have discovered one surprising advantage of trumpet data over 
their wormhole counterpart, which is that
the mass of a boosted Bowen-York trumpet can be prescribed analytically
by the mass parameter in the conformal-factor ansatz used to
solve the Hamiltonian constraint. This is a great computational
advantage over the wormhole case, where the mass parameter 
must be iterated to produce data that contain black holes with
specific desired masses. This relationship could not however 
be extended to spinning black holes.

The motivation to produce black-hole initial data in trumpet form is 
that this is the topology that is preferred by the gauge conditions that
are used in the moving-puncture method, which is itself the most 
popular method for simulating black-hole binaries. 
Although we do not expect (and did not find) any dramatic differences
in the properties of black-hole simulations between wormhole and
trumpet data, the construction of these data are an important first
step towards ideal initial data for puncture simulations. Such ideal
data will be in the 1+log gauge (or whatever slicing condition 
is ultimately used to evolve the data, one natural alternative being
hyperboloidal slicing conditions~\cite{Ohme:2009gn,Buchman:2009ew}), 
will represent true boosted Schwarzschild
or Kerr black holes, and will be in trumpet form. In this work we
have made the simplest step in this direction, i.e., we have produced
trumpet data, but they are maximally sliced and represent only 
approximations to boosted Schwarzschild and Kerr black holes. 

Efforts in these other directions have already been made. 
Data for superposed Kerr punctures have for example been presented
in~\cite{Hannam:2006zt}, and superposed boosted Schwarzschild 
punctures have been used in~\cite{Shibata:2008rq}; non-conformally-flat
data that attempt to include the GW signal from the earlier inspiral of
the binary have been proposed in~\cite{Kelly:2007uc}. Work has also
been done in producing non-conformally-flat data with 
excision techniques~\cite{Lovelace:2008tw,Lovelace:2008hd}. 
It is also now known how to produce 1+log trumpet puncture data for a
single Schwarzschild black hole~\cite{Hannam:2008sg}. It is 
likely that a combination of all of these approaches will be necessary
to produce the optimal data for puncture simulations.

\acknowledgments

We thank Sergio Dain for discussions related to the extreme 
spinning-Bowen-York limit, and Frank Ohme and Julia Gundermann 
for careful readings of the manuscript.

MH and N\'OM were supported by SFI grant 07/RFP/PHYF148. 
SH was supported in part as a VESF fellow of the EGO, by 
DAAD grant D/07/13385 and grant FPA-2007-60220 from the 
Spanish Ministerio de Educaci\'on y Ciencia.

We thank LRZ (Munich), ICHEC (Dublin) and CESGA 
(Santiago de Compostela) for providing computational resources.
MH thanks the University of the Balearic Islands for hospitality while 
some of this work was carried out.

\bibliography{trumpet}

\end{document}